\documentclass[a4paper,11pt]{article}

\pdfoutput=1 

\usepackage{bbm}

\usepackage{jheppub} 

\usepackage[T1]{fontenc}

\DeclareMathAlphabet{\mathantt}{OT1}{antt}{li}{it}
\DeclareMathAlphabet{\mathpzc}{OT1}{pzc}{m}{it} 

\usepackage{simplewick} 

\usepackage{makecell}	
\usepackage{tabu}		

\usepackage[makeroom]{cancel} 

\usepackage[normalem]{ulem} 

\usepackage{amsthm}

\usepackage[utf8]{inputenc}

\usepackage{amsmath,amsfonts,amsthm,amssymb} 

\newcommand{\p}{\partial}

\newcommand{\e}{\epsilon}

\newcommand{\eq}{&\quad}

\newcommand{\rig}{\right.}
\newcommand{\lef}{\left.}

\newcommand{\mco}{\mathcal{O}}

\preprint{UUITP-16/17}

\title{\boldmath Anomalous Dimensions in the WF O($N$) Model with a Monodromy Line Defect}

\author[a]{Alexander Söderberg}

\affiliation[a]{Department of Physics and Astronomy, Uppsala University, Uppsala, Sweden \\}

\emailAdd{alexander.soderberg@physics.uu.se}

\abstract{Implications of inserting a conformal, monodromy line defect in three dimensional O($N$) models are studied. We consider then the WF O($N$) model, and study the two-point Green's function for bulk-local operators found from both the bulk-defect expansion and Feynman diagrams. This yields the anomalous dimensions for bulk- and defect-local primaries as well as one of the OPE coefficients as $\e$-expansions to the first loop order. As a check on our results, we study the $(\phi^k)^2\phi^j$ operator both using the bulk-defect expansion as well as the equations of motion.}

\begin{document} 
	
\newtheorem{defin}{Definition}
\newtheorem{thm}{Theorem}
\newtheorem{cor}{Corollary}
\newtheorem{pf}{Proof}
\newtheorem{nt}{Note}
\newtheorem{ex}{Example}
\newtheorem{ans}{Ansatz}
\newtheorem{que}{Question}
\newtheorem{ax}{Axiom}

\maketitle

\section{Introduction and Review}

Conformal field theories (CFT) in higher than two dimensions are interesting in several different contexts, e.g. condensed matter physics (three dimensions), particle physics (four dimensions), AdS/CFT correspondence \cite{9711200} and entanglement \cite{0603001}. There has been a lot of development in higher dimensional CFTs\footnote{I.e. theories in more than two dimensions.} since the breakthrough in conformal bootstrap \cite{0807.0004}, where the authors numerically determined an upper bound on the dimensions of leading primaries in the OPE, and after the analytical approaches to the bootstrap program for higher dimensional theories \cite{1212.3616, 1212.4103}, where they studied the large spin behavior of CFTs. The results from \cite{1212.3616, 1212.4103} are generalized in \cite{1611.01500}, where a large spin perturbation theory is developed. This method is later used in \cite{1612.00696}\footnote{We thank Alday for telling us about this development.}. Some notable examples of analytical developments in higher dimensional theories are \cite{1601.01784, 1601.06794, 1603.00387, 1606.09593, 1610.06938, 1611.10060, 1701.04830, 1703.03430, 1703.04830, 1703.05325}, as well as numerical developments \cite{1511.02921, 1511.07108, 1512.00013, 1603.04444, 1605.08087}. More important for this paper, are higher dimensional O($N$) models, which also have had a lot of development lately \cite{1510.07770, 1512.05994, 1602.04928, 1607.07077, 1609.09820, 1610.08472, 1612.00696, 1612.05032, 1701.06997, 1702.03938}. It is interesting to study O($N$) models since they are important for the AdS/CFT correspondence, see \cite{1607.07077} and references therein.

\quad\quad Lately there has been a lot of development in CFTs with a defect, i.e. defect conformal field theories (DCFT), both analytically \cite{1601.02883, 1602.06354, 1607.06155, 1608.05126, 1702.08471} and numerically \cite{1210.4258, 1304.4110, 1502.07217, 1605.04175}. Such theories may be used to explain boundary conditions, magnetic-like impurities in spin systems, Rényi entropy and entanglement, see \cite{0303249, 1502.07217, 1607.06155, 1608.05126, 1611.02485} and references therein. A defect is a subspace in the space of a theory, where new operators and interactions between operators may occur. It is therefore important to distinguish between bulk-local operators, which live in the entire space of the theory, and defect-local operators, which only live on the defect. Using the operator product expansion (OPE), it is possible to write bulk-local and defect-local operators in terms of each other when the bulk-local operators are close to the defect \cite{1310.5078, 1602.06354}. We call these OPEs the bulk-defect as well as defect-bulk expansion. These expansions contain OPE coefficients, that are promoted to tensors (with arbitrary many indices) in theories with a global symmetry, as is the case of O($N$) models. The tensors/coefficients in these expansions do not need to be real-valued, unlike the coefficients in the OPE between two bulk-local operators. We expect the global symmetry of the theory to be broken after insertion of a defect, since in general the latter is only left invariant under some subgroups of the global symmetry group. A conformal defect behaves like a CFT on its own. Meaning, conformal transformations parallel to the defect is preserved, i.e. if a conformal defect of codimension $m$ is inserted into a $d$-dimensional CFT, SO($d-m+1,1$) is left unbroken\footnote{In this paper we use Euclidean signature.}. If the defect is flat or spherical, rotations SO($m$) around the defect is preserved as well. This rotation group will act as a global or internal symmetry of the defect-local operators. So a conformal flat defect will break the SO($d+1,1$) conformal group into SO($m$)$\times$SO($d-m+1,1$). In this case, defect-local operators may carry both SO($m$)- and SO($d-m+1,1$)-spin, while bulk-local operators may carry SO($d+1,1$)-spin. Bulk-local operators are transformed under an element from the global symmetry group as they are transported around a monodromy defect. We may define several different defects using different group elements from the global symmetry group in the monodromy transformation.

\quad\quad In this paper we study the implications of inserting a monodromy line defect into a conformal, three dimensional O($N$) model using the bulk-defect expansion. Inserting this defect will break the conformal SO($4,1$) symmetry into SO($2$)$\times$SO($2,1$). The monodromy action tells us about the SO($2$)-spin of the defect-local operators as well as how the global O($N$) symmetry is broken after the defect is inserted, while symmetry of the residual subgroups of O($N$) tell us what kinds of OPE tensors may exist in the bulk-defect expansion, and thus also restricts what kinds of defect-local operators will live on the defect. We find that the global O($N$) symmetry is broken into two or three subgroups, depending on what group element we use in the monodromy action. Operators that transform in different unbroken subgroups do not mix with each other, and defect-local operators in the bulk-defect expansions will transform under the same subgroup as their corresponding bulk-local operator. The SO($2$)-spin of the defect-local operators will differ depending on what subgroup they transform under. This spin, $s_X$, can be generic, and does not need to be integer or half-integer
\begin{align}
\begin{split}
s_X \in \mathbb{Z} + \upsilon \ , \quad \upsilon\in[0,1) \ . 
\end{split}
\end{align}
 
We denote bulk- and defect-local operators that transform in one of the subgroups that are left unbroken, say O($X$), as $\phi_X^j$ and $\psi_{X}^{j}$, where $\psi_{X}^{j}$ has SO($2$)-spin $s_X$. By studying this O($X$) symmetry we find that only vector operators will appear in the bulk-defect expansion, with OPE tensors of rank zero, i.e. OPE constants (denoted $c_X$), in the bulk-defect expansion. 

\quad\quad The 3D Ising model with a monodromy line defect was studied analytically in \cite{1310.5078}. They started from the Wilson-Fisher (WF) fixed point in $4-\e$ dimensional $\phi^4$ theory and let $\e$ go to one (the defect is always of co-dimension two). The scaling dimensions of bulk- and defect-local primaries as well as some of the OPE coefficients were found to the first loop order through comparison of the two-point Green's functions for two bulk-local operators on the defect found in two different ways. One being from the bulk-defect expansion, the other from Feynman diagrams. Their results are in agreement with the numerical data from \cite{1304.4110}. We will generalize this approach to an O($N$) model by promoting the scalar operators in $\phi^4$-theory into vector multiplets of O($N$). We call this theory the WF O($N$) model. The CFT data we find through this approach are\footnote{Here $\tilde{\psi}(x)$ is the digamma function.}
\begin{align} \label{Result}
\begin{split}
|c_X| &= 1 - \frac{\tilde{\psi}(|s_X| + 1) - \tilde{\psi}(1)}{4}\e + \mco(\e^2) \ , \\
\Delta_{\psi_X} &= |s_X| + 1 - \left(1 - \frac{\upsilon(\upsilon - 1)(X + 2)}{(X + 8)|s_X|}\right)\frac{\e}{2} + \mco(\e^2) \ , \\
\Delta_{\phi_X} &= 1 - \frac{\e}{2} + \mco(\e^2) \ .
\end{split}
\end{align}

\quad\quad Another analytical approach is the $\e$-expansion for the 3D Ising model created by Rychkov and Tan in 2015 \cite{1505.00963}. This approach (we will call it the Rychkov-Tan analysis) constrains the theory by defining three axioms that contain information about its dynamics. One of these axioms states that every $\phi^n \ , n \geq 0 \ , n\in\mathbb{Z}$ is a primary, except $\phi^3$ which is a descendant of $\phi$. This follows from the equations of motion. The Rychkov-Tan analysis has been applied to several different theories, e.g. scalar theories in different dimensions \cite{1506.06616, 1605.08868, 1612.08115}, the Gross-Neveu model \cite{1510.04887, 1510.05287}, O($N$) models \cite{1601.01310}, theories studied in Mellin space \cite{1609.00572, 1611.08407}, the Lee-Yang model \cite{1611.06373}, generalized free CFTs \cite{1611.10344} and the 3D Ising model with a monodromy line defect \cite{1607.05551}. The same scaling dimension of defect-local operators as those from \cite{1310.5078} was found using the Rychkov-Tan analysis in \cite{1607.05551}. At the end of this paper we generalize the Rychkov-Tan analysis in \cite{1607.05551} to the WF O($N$) model. We find that the anomalous dimensions for bulk- and defect-local operators are in agreement with the corresponding ones found using the approach in \cite{1310.5078}, see (\ref{Result}), indicating that they are correct.

\quad\quad This paper is outlined as follows. In section \ref{ChO(N)Phi4Model} we study the implications of inserting a monodromy, line defect into a three dimensional O($N$) model. Here we study constraints on the bulk-defect expansion that arises from the monodromy of the defect and the symmetry of the unbroken subgroups of O($N$) that are left preserved after the defect has been inserted. Some technical details about the monodromy constraint are gathered in appendix \ref{AppEqSysSol}. We generalize the approach in \cite{1310.5078} to the WF O($N$) model in section \ref{ChGreenFcn}. The Green's function for two bulk-local operators are studied using both the bulk-defect expansion and Feynman diagrams (up to one loop level). The results (\ref{Result}) are found in this section. We have placed technicalities about the one-loop Feynman integrals in appendix \ref{AppInt}. Finally in section \ref{ChNoPertThy} we generalize the Rychkov-Tan analysis to the WF O($N$) model with a monodromy, line defect. This section serves as a check that our results from section \ref{ChGreenFcn} are correct.

\section{Monodromy Line Defect in a Three Dimensional O($N$) Model} \label{ChO(N)Phi4Model} 

Let us consider a three dimensional CFT with a global O($N$) symmetry and a monodromy line defect. We expect a breaking of the O($N$) symmetry by inserting this defect. Thus we will consider bulk-local fields that are in a vector representation of the residual symmetry group $G$. A monodromy defect is defined with the action
\begin{align} \label{Monodromy}
\begin{split}
\Phi^j(r,\theta+2\pi,y) = g^j{}_{j'}\Phi^{j'}(r,\theta,y) \ , \quad g^j{}_{j'}\in O(N) \ , \quad j \in \{1,...,N\}  \ .
\end{split}
\end{align}

\noindent Here $r$ and $\theta$ are polar coordinates transverse to the defect, and  $y$ is the coordinate parallel to the defect. This condition means that if we transport $\Phi^j$ around the defect, we get back a transformed operator. The choice of the group element $g^j{}_{j'}$ from O($N$) will define the defect.

\begin{ex}
In the $3D$ Ising model, the global symmetry group is $Z_2$. Thus the monodromy defect in this theory can be defined with either $g = \pm 1$. In this case, $g = 1$ is the trivial case when there is no defect. See \cite{1310.5078} for the implications of $g = -1$.
\end{ex}

\noindent If one of these bulk-local operators is close to the defect, we may write it in terms of defect-local operators using the bulk-defect expansion \cite{9505127}. In a three-dimensional CFT with a codimension two defect, the bulk-defect expansion for the rescaled $\Phi^j$ presented in \cite{1310.5078} is generalized into
\begin{align} \label{BulkExpansion0}
\begin{split}
\Phi^j(r,\theta,y) &= \sum_{\Psi^R_s}C^{j}{}_{R,s}\frac{e^{-is\theta}}{r^{\Delta_\Phi - \Delta_\Psi}}B_{\Delta_\Psi}(r^2,\p_y^2)\Psi^R_s(y) \ , \\
B_{\Delta}(x, y) &= \sum_{m\geq 0}\frac{(-1)^m(\Delta)_m}{m!(2\Delta)_{2m}}x^{m}y^{m} \ , \\ 
C^{j}{}_{R,s}\Psi^R_s &\equiv \left(C^{j}{}_{R,s}\right)_{k_1...k_l}\left(\Psi^R_s\right)^{k_1...k_l}  \ , \quad \Delta_\Psi \equiv \Delta_{\Psi^R_s}(R,s) \ .
\end{split}
\end{align}

\noindent Here we sum over all tensor primaries, $\Psi^R_s$, that lives on the defect. These defect-local operators are in irreducible representations $R$ of $G$, and different defect-local operators may transform in different representations of $G$. In this expansion $C^{j}{}_{R,s}$ is an OPE tensor that transforms as a vector of $G$ when contracted with $\Psi^R_s$, $s$ is the SO($2$)-spin of $\Psi^R_s$, $\Delta_\Psi$ is the scaling dimension of $\Psi^R_s$ and $(x)_m$ is the Pochhammer symbol. Note that both $C^{j}{}_{R,s}$ and $\Delta_\Psi$ depend on $R$ and $s$, i.e. they may differ for each $\Psi^R_s$. We can see that the original SO($d+1,1$) conformal symmetry has been broken into SO($2$)$\times$SO($d-1,1$), where SO($2$) describes rotations around the defect, and SO($2,1$) describes conformal transformations parallel to the defect. This expansion is valid only when $\Phi^j$ is close to the defect. Since SO($2$) is an Abelian group, $s$ will act as a charge under the global SO($2$)$\simeq$U($1$) transformations that $\Psi^R_s$ enjoys
\begin{align}
\begin{split}
\Psi^R_s(y) = e^{is\theta}\Psi^R_s(y) \ .
\end{split}
\end{align}

Note that the continuous parameter $\theta$ in this SO($2$)-transformation is one of the polar coordinates in the CFT bulk. The factor $e^{-is\theta}$ in (\ref{BulkExpansion0}) makes sure that $\Psi^R_s(y)$ can transform globally under SO($2$) without affecting $\Phi^j(r,\theta,y)$. Reality of $\Phi^j$ implies that \cite{1310.5078}
\begin{align}
\begin{split}
\Psi^R_{-s} = \bar{\Psi}^R_s \ .
\end{split}
\end{align}

The first thing we need to ask ourselves is what kinds of defect-local operators may appear in the expansion (\ref{BulkExpansion0}). We may be able to constrain the theory using the definition of a monodromy action (\ref{Monodromy}) as well as the residual symmetry $G$. Since we expect the global O($N$) symmetry to be broken by the monodromy of the defect, we have to study constraints on the dynamics from it first.

\subsection{Monodromy Action Constraint} \label{SectionMonodromyDefect}

By conjugation, an O($N$)-matrix is given by\footnote{We can think of this as a general O($N$) transformation where we have chosen the basis vectors in this O($N$) space such that it only rotates the first two vectors.}
\begin{align} \label{UsedDefectDef}
\begin{split}
(g^j{}_{j'})(\vartheta) = \begin{bmatrix}
R_\vartheta & 0 & 0 \\
0 & \mathbbm{1}_{\chi\times\chi} & 0 \\
0 & 0 & -\mathbbm{1}_{(N-\chi-2)\times(N-\chi-2)}
\end{bmatrix} \ , \quad R_\vartheta = \begin{bmatrix}
\pm\cos\vartheta & \mp\sin\vartheta \\
\sin\vartheta & \cos\vartheta
\end{bmatrix} 
\end{split}
\ .
\end{align}

\noindent Here $\chi\in\{0,1,...,N-2\}$. Monodromy of the defect (\ref{Monodromy}) together with the bulk-defect expansion (\ref{BulkExpansion0}) yields
\begin{align} \label{EquationSystem}
\begin{split}
\left\{ \begin{array}{l l}
e^{-2\pi is}C^1{}_{R,s}\Psi^R_s &= \pm\cos\vartheta C^1{}_{R,s}\Psi^R_s \mp\sin\vartheta C^2{}_{R,s}\Psi^R_s \ , \\
e^{-2\pi is}C^2{}_{R,s}\Psi^R_s &= \sin\vartheta C^1{}_{R,s}\Psi^R_s + \cos\vartheta C^2{}_{R,s}\Psi^R_s \ , \\
e^{-2\pi is}C^q{}_{R,s}\Psi^R_s &= C^q{}_{R,s}\Psi^R_s \ , \quad q\in\{3,...,\chi + 2\} \ , \\
e^{-2\pi is}C^r{}_{R,s}\Psi^R_s &= -C^r{}_{R,s}\Psi^R_s \ , \quad r\in\{\chi + 3,...,N\} \ .
\end{array} \right.
\end{split}
\end{align}

\noindent There are two important special cases for the above equation system. These special cases occur when we cannot write $C^1{}_{R,s}\Psi^R_s$ in terms of $C^2{}_{R,s}\Psi^R_s$ and vice versa, i.e. when
\begin{align} \label{SinTheta=0}
\begin{split}
\sin\vartheta = 0 \quad\Leftrightarrow\quad \vartheta = \left\{ \begin{array}{l l}
0 \mod 2\pi \ , \\			
\pi \mod 2\pi \ .			
\end{array} \right.			
\end{split}
\end{align}

\noindent We will get two different sets of solutions depending on whether $R_\vartheta$ describes a proper ($\det R_\vartheta = 1$) or improper ($\det R_\vartheta = -1$) rotation.

\subsubsection{Proper Rotation}

We consider first the two special cases (\ref{SinTheta=0}). If $\vartheta$ equals zero, (\ref{EquationSystem}) reduces to
\begin{align}
\begin{split}
\left\{ \begin{array}{l l}
e^{-2\pi is}C^p{}_{R,s}\Psi^R_s &= C^p{}_{R,s}\Psi^R_s \ , \quad p\in\{1,...,\chi + 2\} \ , \\
e^{-2\pi is}C^r{}_{R,s}\Psi^R_s &= -C^r{}_{R,s}\Psi^R_s \ , \quad r\in\{\chi + 3,...,N\} \ .
\end{array} \right.
\end{split}
\end{align}

\noindent This system has two solutions. Either
\begin{align} \label{Sol1}
\begin{split}
C^r{}_{R,s}\Psi^R_s = 0 \ \ \forall \ r\in\{\chi + 3,...,N\} \ , \quad s = n \ ,
\end{split}
\end{align}

\noindent where $C^p{}_{R,s}\Psi^R_s \ , \ p\in\{1,...,\chi + 2\} \ ,$ does not receive any constraints, or 
\begin{align} \label{Sol2}
\begin{split}
C^p{}_{R,s}\Psi^R_s = 0 \ \ \forall \ p\in\{1,...,\chi + 2\} \ , \quad s = n + \frac{1}{2} \ ,
\end{split}
\end{align}

\noindent where $C^r{}_{R,s}\Psi^R_s \ , \ r\in\{\chi + 3,...,N\} $ does not receive any constraints. In this section $n$ is an integer, i.e. $n \in\mathbb{Z}$. The solutions (\ref{Sol1}) and (\ref{Sol2}) tell us that the global O($N$) symmetry group has been broken into 
\begin{align}
\begin{split}
G = \text{O}(\chi + 2)\times O(N-\chi - 2) \ .
\end{split}
\end{align}

The branching rule tells us that $\Phi^j$ can be separated into bulk-local operators that transform in O($\chi + 2$) and bulk-local operators that transform in O($N-\chi - 2$)
\begin{align}
\begin{split}
\Phi^j = \phi_{\chi + 2}^a \oplus \phi_{N - \chi - 2}^b \ , \quad a\in\{1,...,\chi + 2\} \ , \quad b\in\{1,...,N - \chi - 2\} \ .
\end{split}
\end{align}

\noindent Both $\phi_{\chi + 2}^a$ and $\phi_{N - \chi - 2}^b$ will have bulk-defect expansions similar to (\ref{BulkExpansion0}). The defect-local operators in these expansions will transform under the same orthogonal symmetry group as their corresponding bulk-local operator, e.g. the defect-local operators, $\psi_{\chi + 2}$, in the bulk-defect expansion of $\phi_{\chi+2}^a$ will transform under O($\chi + 2$). The SO($2$)-spin of $\psi_{\chi + 2}$ will be an integer, while the SO($2$)-spin of $\psi_{N - \chi - 2}$ will be a half-integer spin, i.e.
\begin{align}
\begin{split}
\quad s_{\chi + 2} = n \ , \quad s_{N - \chi - 2} = n + \frac{1}{2} \ .
\end{split}
\end{align}

\noindent More precisely, we can write the bulk-defect expansion (\ref{BulkExpansion0}) for the original bulk-local operator (that transforms in $G$) as two sums. One that sums over defect-local primaries with integer spins, and one that sums over defect-local primaries with half-integer spins. The first of these sums corresponds to $\phi^a_{\chi+2}$ and contains $\psi_{\chi+2}$, while the second corresponds to $\phi^b_{N-\chi-2}$ and contains $\psi_{N-\chi-2}$. A similar decomposition is possible for all of the other cases studied in this section as well.
\begin{align}
\begin{split}
\Phi^j(r,\theta,y) &= \sum_{\psi_{\chi + 2}}C^{j}_{\chi + 2}\frac{e^{-is_{\chi + 2}\theta}}{r^{\Delta_{\phi_{\chi+2}} - \Delta_{\psi_{\chi+2}}}}B_{\Delta_{\psi_{\chi+2}}}\psi_{\chi + 2} + \\
&\quad+ \sum_{\psi_{N-\chi - 2}}C^{j}_{N - \chi - 2}\frac{e^{-is_{N - \chi - 2}\theta}}{r^{\Delta_{\phi_{N-\chi-2}} - \Delta_{\psi_{N-\chi-2}}}}B_{\Delta_{\psi_{N-\chi - 2}}}\psi_{N-\chi - 2} \ , \\
C^j_X\psi_X &\equiv (C^j_X)_{k_1...k_l}(\psi_X)^{k_1...k_l} \ , \\
\Delta_{\psi_X} &\equiv \Delta_{\psi_X}(R_X, s_X) \ , \quad C^j_X \equiv C^j_X(R_X, s_X) \ .
\end{split}
\end{align}

Here $X\in \{\chi + 2, N- \chi - 2\}$ and $\psi_X$ is a defect-local primary that is in the irreducible representation $R_X$ of O($X$), and transforms as a vector of O($X$) when contracted with the OPE coefficient $C^j_X$. It has SO($2$)-spin $s_X$ and scaling dimension $\Delta_{\psi_X}$. The OPE coefficients, the scaling dimensions (for both the bulk-local and defect-local operators) as well as the irreducible representations of the defect-local operators in these two sums may be different to each other. \\

\noindent It is a similar story when $\vartheta = \pi$. The O($N$) symmetry is then broken into O($\chi$) $\times$ O($N - \chi$), and defect-local operators that transform in O($\chi$) have integer SO($2$)-spin, while defect-local operators that transform in O($N - \chi$) have half-integer SO($2$)-spin. \\

\noindent A more interesting case is when we consider $\vartheta$ to be generic, i.e. $\sin\vartheta \ne 0$. Then (\ref{EquationSystem}) yields the following system of equations\footnote{See the "Proper Rotation" section of appendix \ref{AppEqSysSol} for details on this. In this paper we assume that the OPE tensors can be complex-valued.}
\begin{align} \label{ProperEqSys}
\begin{split}
\left\{ \begin{array}{l l}
C^1{}_{R,s}\Psi^R_s &= \pm iC^2{}_{R,s}\Psi^R_s \ , \quad s = n + \frac{\vartheta}{2\pi} \ , \quad n\in\mathbb{Z} \ , \\
e^{-2\pi is}C^q{}_{R,s}\Psi^R_s &= C^q{}_{R,s}\Psi^R_s \ , \quad q\in\{3,...,\chi + 2\} \ , \quad s = n' \ , \quad n'\in\mathbb{Z} \ , \\
e^{-2\pi is}C^r{}_{R,s}\Psi^R_s &= -C^r{}_{R,s}\Psi^R_s \ , \quad r\in\{\chi + 3,...,N\} \ , \quad s = n'' + \frac{1}{2} \ , \quad n''\in\mathbb{Z} \ .
\end{array} \right.
\end{split}
\end{align}

\noindent These constraints are on the dynamics of the theory coming from the monodromy action. We see that the first two components of $C^j{}_{R,s}\Psi^R_s$ relate to each other, and do not mix with other components of the tensor. The system of equations (\ref{ProperEqSys}) has three solutions\footnote{The solutions can be read off by matching the spin required for the equations to hold.}. Either
\begin{align} \label{O(2)DefectLocals}
\begin{split}
C^1{}_{R,s}\Psi^R_s = \pm iC^2{}_{R,s}\Psi^R_s \ , \quad C^v{}_{R,s}\Psi^R_s = 0 \ \ \forall \ v\in\{3,...,N\} \ , \quad s = n + \frac{\vartheta}{2\pi} \ ,
\end{split}
\end{align}

\noindent or
\begin{align}
\begin{split}
C^{v'}{}_{R,s}\Psi^R_s = 0 \ \ \forall \ v'\in\{1,2,\chi + 3,...,N\} \ , \quad s = n \ ,
\end{split}
\end{align}

\noindent where $C^q{}_{R,s}\Psi^R_s \ , \ q\in\{3,...,\chi + 2\}$ does not receive any constraints, or
\begin{align}
\begin{split}
C^{v''}{}_{R,s}\Psi^R_s = 0 \ \ \forall \ v''\in\{1,...,\chi + 2\} \ , \quad s = n + \frac{1}{2} \ ,
\end{split}
\end{align}

\noindent where $C^r{}_{R,s}\Psi^R_s \ , \ r\in\{\chi + 3,...,N\}$ does not receive any constraints. Thus the O($N$) symmetry has been broken into O($2$) $\times$ O($\chi$) $\times$ O($N - \chi - 2$), where defect-local operators that transform under O($2$) have generic SO($2$)-spin, defect-local operators that transform under O($\chi$) have integer SO($2$)-spin and defect-local operators that transform under \\
O($N - \chi - 2$) have half-integer SO($2$)-spin. Note that the two components of the bulk-defect expansion for the bulk-local operator, $(\phi^a_2) = (\Phi^1, \Phi^2)$, that transforms in O($2$) are related through (\ref{O(2)DefectLocals}).

\begin{nt}
If we consider an O($N$)-model where the OPE tensors need to be real, the relation (\ref{O(2)DefectLocals}) yields that $C^1{}_{R,s}\Psi^R_s$ and $C^2{}_{R,s}\Psi^R_s$ are zero and thus also $\Phi^1$ and $\Phi^2$ are zero. In this case the O($N$) symmetry is broken into O($\chi$)$\times$O($N-\chi-2$).
\end{nt}

\subsubsection{Improper Rotation}

The solutions to (\ref{EquationSystem}) considering the special cases when $\vartheta$ equals zero or $\pi$ will yield similar solutions as those in the proper case. In both of these cases the global O($N$) symmetry is broken, leaving a O($\chi + 1$) $\times$ O($N - \chi - 1$) symmetry. Defect-local operators that transform in O($\chi + 1$) will have integer SO($2$)-spin, while defect-local operators that transform in O($N - \chi - 1$) will have half-integer SO($2$)-spin. The procedure of finding this is exactly the same as that discussed in the previous section. \\

\noindent If we consider a generic angle, i.e. $\sin\vartheta \ne 0$, the results will differ from the proper case. The system of equations (\ref{EquationSystem}) yields\footnote{See the "Improper Rotation" section of appendix \ref{AppEqSysSol} for details on this.}
\begin{align} \label{ImproperEqs}
\begin{split}
\left\{ \begin{array}{l l}
C^1{}_{R,s}\Psi^R_s &= \frac{\sin(\vartheta)}{e^{-2\pi is} + \cos(\vartheta)}C^2{}_{R,s}\Psi^R_s \ , \quad s = \frac{n}{2} \ , \quad n\in\mathbb{Z} \ , \\
e^{-2\pi is}C^q{}_{R,s}\Psi^R_s &= C^q{}_{R,s}\Psi^R_s \ , \quad q\in\{3,...,\chi + 2\} \ , \quad s = n' \ , \quad n'\in\mathbb{Z} \ , \\
e^{-2\pi is}C^r{}_{R,s}\Psi^R_s &= -C^r{}_{R,s}\Psi^R_s \ , \quad r\in\{\chi + 3,...,N\} \ , \quad s = n'' + \frac{1}{2} \ , \quad n''\in\mathbb{Z} \ .
\end{array} \right.
\end{split}
\end{align}

\noindent As in the proper case, these are constraints on the OPE tensors coming from the monodromy action. Here it is convenient to define two linear combinations of $C^1{}_{R,s}\Psi^R_s$ and $C^2{}_{R,s}\Psi^R_s$
\begin{align}
\begin{split}
\tilde{C}^1{}_{R,s}\Psi^R_s &\equiv C^1{}_{R,s}\Psi^R_s + \frac{\sin\vartheta}{1 - \cos\vartheta}C^2{}_{R,s}\Psi^R_s \ , \\
\tilde{C}^2{}_{R,s}\Psi^R_s &\equiv C^1{}_{R,s}\Psi^R_s - \frac{\sin\vartheta}{1 + \cos\vartheta}C^2{}_{R,s}\Psi^R_s \ .
\end{split}
\end{align}

The system of equations (\ref{ImproperEqs}) have two solutions. Either
\begin{align} \label{IntegerSpinFields}
\begin{split}
\tilde{C}^2{}_{R,s}\Psi^R_s &= 0 \ , \quad C^r{}_{R,s}\Psi^R_s = 0 \ \ \forall \ r\in\{\chi + 3,...,N\} \ , \quad s = n \ ,
\end{split}
\end{align}

\noindent where $\tilde{C}^1{}_{R,s}\Psi^R_s$ and $C^q{}_{R,s}\Psi^R_s \ , \ q\in\{3,...,\chi + 2\}$ does not receive any constraints, or
\begin{align} \label{HalfIntegerSpinFields}
\begin{split}
\tilde{C}^1{}_{R,s}\Psi^R_s = 0 \ , \quad C^q{}_{R,s}\Psi^R_s = 0 \ \ \forall \ q\in\{3,...,\chi + 2\} \ , \quad s = n + \frac{1}{2} \ .
\end{split}
\end{align}

\noindent where $\tilde{C}^2{}_{R,s}\Psi^R_s$ and $C^r{}_{R,s}\Psi^R_s \ , \ r\in\{\chi + 3,...,N\}$ does not receive any constraints. These solutions tells us that the symmetry group has again been broken into \\
O($\chi + 1$) $\times$ O($N - \chi - 1$), where defect-local operators that transform in O($\chi + 1$) have integer SO($2$)-spin, while defect-local operators that transform in O($N-\chi-1$) have half-integer SO($2$)-spin. The linear combination $\tilde{C}^1{}_{R,s}\Psi^R_s$ will appear in the bulk-defect expansion of $\phi^a_{\chi + 1}$, and $\tilde{C}^2{}_{R,s}\Psi^R_s$ will appear in the bulk-defect expansion of $\phi^b_{N - \chi - 1}$. We can check that this result is correct by representing the $C^j{}_{R,s}\Psi^R_s$-terms in the bulk-defect expansions of bulk-local operators that transform in O($\chi + 1$) and O($N - \chi - 1$) as vectors, $\sigma_{\chi + 1}$ and $\sigma_{N - \chi - 1}$, both containing $N$ elements. These elements are the coefficients in front of $C^1{}_{R,s}\Psi^R_s$, ..., $C^N{}_{R,s}\Psi^R_s$, i.e.
\begin{align}
\begin{split}
\sigma_{\chi + 1} &\equiv (C^1{}_{R,s}, ..., C^N{}_{R,s}) = (1,(1 - \cos\vartheta)^{-1}\sin\vartheta,\underset{\chi}{\underbrace{1,...,1}},\underset{N - \chi - 2}{\underbrace{0,...,0}}) \ , \\
\sigma_{N - \chi - 1} &\equiv (C^1{}_{R,s}, ..., C^N{}_{R,s}) = (1,-(1 + \cos\vartheta)^{-1}\sin\vartheta,\underset{\chi}{\underbrace{0,...,0}},\underset{N - \chi - 2}{\underbrace{1,...,1}}) \ .
\end{split}
\end{align}

\noindent Since operators that transform in O($\chi + 1$) should not mix with operators that transform in O($N - \chi - 1$), the two vectors $\sigma_{\chi + 1}$ and $\sigma_{N - \chi - 1}$ should be orthogonal to each other. Indeed, using the trigonometric identity we see that this is the case. Moreover, these two vectors should be eigenvectors to improper O($N$) matrix (\ref{UsedDefectDef}). One can check that this is the case, where $\sigma_{\chi+1}$ has eigenvalue $+1$, and $\sigma_{\chi+1}$ has eigenvalue $-1$. Actually, eigenvectors of improper $g^j{}_{j'}(\vartheta)$ can only obtain the two different eigenvalues $\pm 1$, indicating that our results are correct. \\

\noindent Putting it all together, inserting a monodromy defect using a proper O($2$) rotation, i.e. \\
$\det R_\vartheta = 1$, possibly (depending on the angle $\vartheta$) breaks the global O($N$) symmetry into three parts O($2$) $\times$ O($\chi$) $\times$ O($N - \chi - 2$), where operators that transform in one of these subgroups does not mix with operators from the other subgroups. Each of these bulk-local operators will have a bulk-defect expansion with defect-local operators that transform under the same unbroken subgroup as their corresponding bulk-local operator. The defect-local operators will have different SO($2$)-spins depending on what subgroup they transform under. The situation is very similar when considering an improper O($2$) rotation, i.e. $\det R_\vartheta = - 1$, when defining the defect. In this case however, the global O($N$) symmetry (independently of the angle $\vartheta$) breaks into O($\chi + 1$) $\times$ O($N - \chi - 1$), meaning that in general, using $\det R_\vartheta = - 1$ does not break the symmetry as much as when using  $\det R_\vartheta = 1$.

\begin{nt}
Similar to \cite{1310.5078}, the monodromy action constrains the spin of defect-local operators.
\end{nt}

\noindent The theory is consistent with flipping the defect, i.e. the discussion in this section is the same when we use the following monodromy action
\begin{align} \label{FlippinTheBean}
\begin{split}
\Phi^j(r,\theta - 2\pi,y) = (g^j{}_{j'})^{-1}\Phi^{j'}(r,\theta,y) \ , \quad g^j{}_{j'}\in\mathcal{G} \ .
\end{split}
\end{align}

\subsection{Symmetry Constraints}

In this section we study constraints from the broken O($N$) symmetry. The transformed bulk-local operator, $\phi_X^j$, is to be the same as when we transform the defect-local operators, $\psi^{k_1...k_l}_{X}$, inside the bulk-defect expansion (\ref{BulkExpansion0}). Let $\Omega^j{}_k\in$O($X$) be a transformation matrix from one of the subgroups that is preserved after the global O($N$) symmetry has been broken. Then the transformation of $\phi_X^j$ under $\Omega^j{}_k$ must be compatible with the transformation of $\psi^{k_1...k_l}_{X}$ under the same $\Omega^j{}_k$
\begin{align} \label{TwoSides}
\begin{split}
\Omega^j{}_{j'}\phi_X^{j'} &= \sum_{\psi_{X}^{k_1...k_l}}(C^j_X)_{k_1'...k_l'}\frac{e^{-is_X\theta}}{r^{\Delta_{\phi_X} - \Delta_{\psi_{X}}}}B_{\Delta_{\psi_X}}(r,\p_y)\prod_{n=1}^{l}\Omega^{k_n'}{}_{k_n}(\psi_X)^{k_1...k_l} \ .
\end{split}
\end{align}

Comparing the two sides of this constrains the OPE tensors. It tells us that $(C^j_X)_{k_1...k_l}$ is an isotropic tensor (or tensor invariant) of O($X$)\footnote{Remember that the inverse of an O($X$) matrix is its own transpose.}
\begin{align} \label{TensorInvOfO(N)}
\begin{split}
(C^j_X)_{k_1...k_l} = \Omega_{j'}{}^j(C^{j'}_X)_{k_1'...k_l'}\prod_{n=1}^{l}\Omega^{k_n'}{}_{k_n} \ .
\end{split}
\end{align}

Since there are no vector invariants of O($X$), there cannot be any scalars on the defect.

\quad\quad A general isotropic tensor of O($X$) is given by a sum over all possible permutations of Kronecker deltas \cite{BookGroups}\footnote{These tensors can also be written as a sum over products of Kronecker deltas, even number of Levi-Civitas and combinations of those two. However, even numbers of Levi-Civitas can be written as a product of several Kronecker deltas. Terms with uneven numbers of Levi-Civitas are not invariants of O($X$), but of the smaller group SO($X$). We thank Jian Qiu for telling us about this.}. Since the defect-local operators are in irreducible representations of O($X$), and since tensors with two or more indices in such representations are traceless\footnote{Meaning that if we contract any two of one of these tensors' indices with eachother, it is zero.}, we end up with only vector-operators on the defect
\begin{align} \label{BulkExpansion}
\begin{split}
\phi_X^j(r,\theta,y) &= \sum_{\psi^{j}_{X}}c_X(s_X)\frac{e^{-is_X\theta}}{r^{\Delta_{\phi_X} - \Delta_{\psi_{X}}}}B_{\Delta_{\psi_X}}(r,\p_y)\psi^{j}_{X}(y) \ .
\end{split}
\end{align}

Here $c_X$ is an OPE coefficient and $\psi_X^j$ is an O($X$) vector primary on the defect. Note that it only exist one vector representation of O($X$). \\

In this section we inserted a monodromy line defect into a three dimensional CFT with a global symmetry. From the monodromy action we found how the global symmetry is broken as well as what kinds of SO($2$)-spin the defect-local operators will carry, while from symmetry arguments we found what kinds of defect-local operators can appear in the bulk-defect expansion. All we needed in order to perform this procedure was essentially the bulk-defect expansion (\ref{BulkExpansion0}), which can be used for any three-dimensional CFT with bulk-local vector operators and a codimension two defect. Thus we should be able to apply this procedure to other three-dimensional CFTs with other global symmetries as well. It would be interesting to study bulk-defect expansions in $d$-dimensional CFTs with a monodromy defect of codimension other than two, such that we could perform this procedure to those kinds of theories as well. Note that equation (\ref{TensorInvOfO(N)}) should hold for any symmetry preserved by the defect. Thus OPE tensors in bulk-defect expansions will always be isotropic tensors of the global symmetry group their respective bulk-local operators transform under.

\section{Green's Function} \label{ChGreenFcn}

In this section we generalize the steps in \cite{1310.5078} to the case with O($X$) symmetry. Our starting point for this discussion is Green's function, i.e. the correlator, for two bulk-local operators close to the defect that transform under the same unbroken symmetry group, say O($X$). If the bulk-local operators in this correlator would not be close to the defect, this Green's function would be the usual one we encounter in a CFT without a defect. We proceed to find this Green's function from both the bulk-defect expansion and Feynman diagrams\footnote{Our results from section \ref{SectionMonodromyDefect} tell us that if the bulk-local operators in the two-point correlators transform in same unbroken symmetry group, the SO($2$)-spin in their bulk--defect expansions will be of the same kind, i.e. integer, half-integer or neither.}.
	
\quad\quad The WF O($N$) model is governed by the Lagrangian
\begin{align} \label{Lagrangian}
\begin{split}
\mathcal{L} = \frac{1}{2}(\p_\mu\Phi^j)^2 + \frac{\lambda}{4!}[(\Phi^j)^2]^2 \ , \quad j\in\{1,...,N\}
\end{split}
\ .
\end{align}

We renormalize it using dimensional regularization, i.e. we consider $4 - \epsilon$ dimensions. The $\beta$-function is given by \cite{1404.1094}
\begin{align}
\begin{split}
\beta(\lambda)= \frac{\lambda}{3!}\left(-\e + \frac{N+8}{3!8\pi^2}\lambda\right) + \mathcal{O}(\e^3) \ ,
\end{split}
\end{align}

which have fixed points at
\begin{align} \label{FixedPoint}
\begin{split}
\lambda = 0 \quad \text{and} \quad \lambda = \frac{3!8\pi^2\e}{N + 8} + \mathcal{O}(\e^2) \ .
\end{split}
\end{align}

We consider the CFT at the fixed point where the coupling constant is non-zero.

\subsection{Green's Function from the Bulk-Defect Expansion}

Since we expand (in $\e$) around the free theory, the CFT data will not be degenerate, i.e. there will only be one defect-local operator with SO($2$)-spin $s_X$. Thus we can sum over $s_X$ instead of the defect-local primaries in the bulk-defect expansion (\ref{BulkExpansion}). This yields the full two-point correlator for two bulk-local operators that transforms in O($X$)\footnote{We do not make any assumptions on whether the OPE coefficients, $c_X$, are real-valued.}
\begin{align} \label{PrimalGreenFcn}
\begin{split}
G^{jj'} &\equiv \langle 0|\phi_X^j(r_1,\theta_1,y_1)\phi_X^{j'}(r_2,\theta_2,y_2)|0\rangle \\
&= \sum_{s_X,s'_X}c_X^\dagger c_X\frac{e^{i(s_X\theta_1 - s'_X\theta_2)}}{r_1^{\Delta_{\phi_X} - \Delta_{\psi_X}}r_2^{\Delta_{\phi_X} - \Delta_{\psi'_X}}}\times \\
\eq \times
\left[1 + \mathcal{O}(r_1^2\p_{y_1}^2) + \mathcal{O}(r_2^2\p_{y_2}^2)\right]\langle 0|\psi_{X}^{j}(y_1)\psi_{X}^{\prime j'}(y_2)|0\rangle \ ,
\end{split}
\end{align}

where the SO($2$)-spins, $s_X$ and $s_X'$, will be of the same kind
\begin{align} \label{Spin}
\begin{split}
s_X \ , \ s_X' \in \mathbb{Z} + \upsilon \ , \quad \upsilon\in[0,1) \ . 
\end{split}
\end{align}

Here $\upsilon$ is fixed and the same for both $s_X$ and $s_X'$ since operators with different kinds of SO($2$)-spin do not mix with each other (see section \ref{SectionMonodromyDefect}). The defect-local operators are normalized through its two-point correlator
\begin{align}
\begin{split}
\langle 0|\psi_{X}^{j}(y_1)\psi_{X}^{\prime j'}(y_2)|0\rangle = \frac{\delta^{\psi_{X}^{j}\psi_{X}^{\prime j'}}}{|y_{12}|^{2\Delta_{\psi_X}}} \ , \quad \delta^{\psi_{X}^{j}\psi_{X}^{\prime j'}} = \delta_{s_Xs'_X}\delta^{jj'}  \ , \quad y_{12} \equiv y_1 - y_2 \ . 
\end{split}
\end{align}

We place the bulk-local operators on the same distance from the defect, i.e. $r \equiv r_1 = r_2$
\begin{align} \label{Green'sFcnOPE}
\begin{split}
G^{jj'}_{s_X} = |c_X|^2\frac{e^{is_X\theta_{12}}}{r^{2\Delta_{\phi_X}}}\rho^{2\Delta_{\psi_X}}\delta^{jj'}\left[1 + \mathcal{O}(\rho^2)\right] \ , \quad \theta_{12} \equiv \theta_1 - \theta_2 \ , \quad \rho \equiv \frac{r}{|y_{12}|} \ . 
\end{split}
\end{align}

Here $G_{s_X}^{jj'}$ is the summand of (\ref{PrimalGreenFcn}). By comparing this OPE with the result that we will calculate from diagrams at tree-level, we find the zeroth loop order correction to $\Delta_{\phi_X} \ , \Delta_{\psi_X}$ and $|c_X|$. The logarithm of $G^{jj'}_{s_X}$ will be useful when finding correction from one-loop diagrams
\begin{align} \label{LogOPE}
\begin{split}
\log G^{jj'}_{s_X} &= \left(2\log|c_X| + is_X\theta_{12} - 2\Delta_{\phi_X}\log r + 2\Delta_{\psi_X}\log\rho\right)\delta^{jj'} + \mco(\rho^2) \ .
\end{split}
\end{align}

\begin{nt}
	Since bulk-local operators will not be affected by the defect if they are far away from it, we expect the CFT data (in our case $\Delta_{\phi_X}$) for those kind of operators to be the same as the theory without a defect. 
\end{nt}

\subsection{Green's Function from Feynman Rules}

When calculating diagrams using Feynman rules, we calculate one loop order at a time, hence we write Green's function as a sum over loop order corrections, where $G_n$ represents the correction from the $n^\text{th}$ loop order
\begin{align} \label{Green'sFcnExp}
\begin{split}
G^{jj'} = \sum_{n\geq 0}G_n^{jj'} \ .
\end{split}
\end{align}

The logarithm of $G^{jj'}$ will be useful when finding first loop order corrections to the CFT data. We Taylor expand the logarithm of the above sum so it later can be compared with the result from the OPE (\ref{LogOPE})
\begin{align} \label{LogGreen'sFcnExp}
\begin{split}
\log G^{jj'} &= \log G^{jj'}_0 + \left(G_0^{-1}\right)^{j}{}_{j''}G_1^{j''j'} + \mathcal{O}(\epsilon^2) \ .
\end{split}
\end{align}

\subsubsection{Tree-Level Diagram}

The calculation of the tree-level diagram is the same as in \cite{1310.5078}, but with an overall factor of $\delta^{jj'}$ as well as different spin in the spectrum of defect-local operators. These calculations will be expressed in terms of the dimension of the defect
\begin{align}
\begin{split} 
D = 2 - \e \ .
\end{split}
\end{align}

Our starting point is the Laplace equation for the two-point correlator
\begin{align} \label{LaplaceEq}
\begin{split} 
-\nabla^2G_{0}^{jj'}(x_1, x_2) &= \frac{4\pi^{D/2 + 1}}{\Gamma(D/2)}\delta^{jj'}\delta^D(x_1-x_2) \ .
\end{split}
\end{align}

\begin{nt}
This Green's function is normalized such that it has the asymptotic
\begin{align} \label{Normalization}
\begin{split}
G_{0}^{jj'}(x_1, x_2) \equiv \langle 0|\phi_X^j(x_1)\phi_X^{j'}(x_2)|0\rangle &= \frac{\delta^{jj'}}{|x_1-x_2|^D} \ .
\end{split}
\end{align}
\end{nt}

In momentum space, the Laplace equation (\ref{LaplaceEq}) has the solution
\begin{align}
\begin{split}
G_{0}^{jj'}(x_1, x_2) &= \frac{2\pi^{D/2}}{\Gamma(D/2)}\delta^{jj'}\sum_{s_X}\int\frac{d^Dk}{(2\pi)^D}e^{is_X\theta_{12}}e^{iky_{12}}I_{|s_X|}(kr_-)K_{|s_X|}(kr_+) \ , \\
r_- &= \min(r_1,r_2) \ , \quad r_+=\max(r_1,r_2) \ .
\end{split}
\end{align}

Here $I_{|s_X|}$ and $K_{|s_X|}$ are modified Bessel functions. Using some relations that the modified Bessel functions satisfy, we can rewrite the summand, $G_{0s_X}^{jj'}$, of $G_0^{jj'}$ as
\begin{align} \label{PrimalZeroLoopGreen'sFcn} 
\begin{split}
G_{0s_X}^{jj'}(x_1, x_2) &= \frac{\Gamma(|s_X| + D/2)}{\Gamma(D/2)\Gamma(|s_X| + 1)}\frac{e^{is_X\theta_{12}}}{\left(r_1r_2\right)^{D/2}}(4\xi)^{-(|s_X| + D/2)}\delta^{jj'}\times \\
\eq \times{}_2F_1\left(|s_X| + D/2, |s_X| + 1/2, 2|s_X| + 1, -\xi^{-1}\right) \ , \\
\xi &= \frac{y_{12}^2 + r_{12}^2}{4r_1r_2} \ , \quad r_{12} = r_1 - r_2 \ .
\end{split}
\end{align}

Here ${}_2F_1$ is a hyper geometric function, and $\xi$ is one of the two conformally invariant cross-ratios. The other one being the relative angle, $\theta_{12}$, between the two bulk-local operators with respect to the defect \cite{1601.02883}. We place the bulk-local operators on the same distance from the defect, i.e. $r \equiv r_1 = r_2$, so we can compare it with the result from the OPE
\begin{align} \label{Green'sFCNSummand}
\begin{split}
G_{0s_X}^{jj'}(x_1, x_2) &= \frac{\Gamma(|s_X| + D/2)}{\Gamma(D/2)\Gamma(|s_X| + 1)}\frac{e^{is_X\theta_{12}}}{r^D}\rho^{2|s_X| + D}\delta^{jj'}\left[1 + \mathcal{O}(\rho^2)\right] \ .
\end{split}
\end{align}

Comparing this with (\ref{Green'sFcnOPE}) yields\footnote{Here we have Taylor expanded $|c_X|_0$ around $\e = 0$.}
\begin{align}
\begin{split}
|c_X|_0 &= 1 - \frac{\tilde{\psi}(|s_X| + 1) - \tilde{\psi}(1)}{4}\e + \mco(\e^2) \ , \\
\Delta_{\phi_X}^0 &= 1 - \frac{\e}{2} \ , \quad \Delta_{\psi_X}^0 = |s_X| + 1 - \frac{\e}{2} \ .
\end{split}
\end{align}

Here $\tilde{\psi}(x)$ is the digamma function, $|c_X|_m$ is the $m$-loop correction to $|c_X|$, and $\Delta_\phi^m/\Delta_\psi^m$ is the $m$-loop correction to $\Delta_{\phi_X}/\Delta_{\psi_X}$. We do not have any constraints on whether OPE coefficients in bulk-defect expansions are real.

\begin{nt}
It is important to remember that in all of the $\epsilon$-expansions in this section, $\epsilon$ is not small, but one. Taking $\epsilon$ to one is not strictly speaking justified, but it is common practice, and it is known to be a good approximation, e.g. the results in the $\phi^4$ case seems to be true by comparing it to numerical data \cite{1304.4110, 1310.5078}.
\end{nt}

\subsubsection{One-Loop Diagram}

The two-point, one-loop diagram (not in momentum space) for bulk-local operators on the defect is given by
\begin{align} \label{OneLoopGreen'sFcn}
\begin{split}
G_{1}^{jj'}(x_1, x_2) &= \sum_{s_X} G_{1s_X}^{jj'}(x_1, x_2) \ , \\
G_{1s_X}^{jj'}(x_1, x_2) &= -\frac{2\lambda}{(2\pi)^4S}\int_{\mathbb{R}^4}d^4x_0 \left( G_{0s_X}^{jk}(x_1, x_0) G_{0kl}(x_0, x_0) G_{0s_X}^{lj'}(x_0, x_2) + \rig \\
\eq + G_{0s_X}^{jk}(x_1, x_0) G_{0lk}(x_0, x_0) G_{0s_X}^{lj'}(x_0, x_2) + \\
\eq\lef + G_{0s_X}^{jk}(x_1, x_0) G_{0l}^l(x_0, x_0) G_{0s_Xk}{}^{j'}(x_0, x_2) \right) \ , \\
S &= 3!2 \ .
\end{split}
\end{align}

Here $\lambda$ is the coupling constant at the WF fixed point, see (\ref{FixedPoint}), and $S$ is the symmetry factor. Please note that in each of the terms in $G_{1s_X}^{jj'}$, one of the Green's functions, $G_{0}^{jj''}$, is the whole sum and not only the summand, $G_{0s_X}^{jj''}$, of (\ref{PrimalZeroLoopGreen'sFcn}). In appendix section \ref{AppSummand} we rewrite $G_{0s_X}^{jj''}$ using hypergeometric function relations
\begin{align}
\begin{split}
G_{0s_X}^{jj''}(x_k, x_l) &= e^{is_X\theta_{kl}}\frac{\left(4r_kr_l\right)^{|s_X|}}{d^-_{kl}d^+_{kl}\left(d^-_{kl} + d^+_{kl}\right)^{2|s_X|}}\delta^{jj'} \ , \\
d_{kl}^\pm &= \sqrt{y_{kl}^2 + (r_{kl}^\pm)^2 + z_{kl}^2} \ , \quad r_{kl}^\pm = r_k \pm r_l \ . 
\end{split}
\end{align}

The sum $G_{0}^{jj''}$ is the propagator for the theory. Renormalization yields that we only need to care about the finite piece of this propagator when we perform the resummation\footnote{Details about this resummation is in appendix section \ref{AppResum}.}
\begin{align} \label{RealProp}
\begin{split}
G_{0}^{jj'}(x_0, x_0) = \frac{\upsilon(\upsilon - 1)}{2r_0^2}\delta^{jj'} \ .
\end{split}
\end{align}

Inserting $G_{0s_X}^{jj'}$ and $G_0^{jj'}$ back into (\ref{OneLoopGreen'sFcn}) yields
\begin{align}
\begin{split}
G_{1s_X}^{jj'}(x_1, x_2) &= -\frac{\upsilon(\upsilon - 1)\lambda}{(2\pi)^4S}e^{is_X\theta_{12}}\left(2 + \delta^l{}_l\right)\delta^{jj'} \times \\
\eq \times \int_\kappa\frac{dy_0dz_0r_0dr_0d\theta_0}{r_0^2}\frac{\left(4rr_0\right)^{2|s_X|}}{d_{10}^-d_{10}^+\left(d_{10}^- + d_{10}^+\right)^{2|s_X|}d_{02}^-d_{02}^+\left(d_{02}^- + d_{02}^+\right)^{2|s_X|}} \ , \\
\kappa &= \{y_0,z_0 \in \mathbb{R} \ , \quad r_0 \in \{0,\infty\} \ , \quad \theta_0 \in \{0,2\pi\} \} \ . 
\end{split}
\end{align}

Here we are using cylindrical coordinates and the positions $x_1$ and $x_2$ are at the same distance from the defect, i.e. $r \equiv r_ 1 = r_ 2$, as well as $z_1 = z_2 = 0$. We rewrite this integral using the variable change
\begin{align} \label{VariableChange2}
\begin{split}
y_0' = y_0 + \frac{y}{2} \ , \quad y \equiv y_{12} \ , 
\end{split}
\end{align}

which yields
\begin{align}
\begin{split}
d_{10}^\pm \stackrel{(\ref{VariableChange2})}{=} \sqrt{\left(y_0' - \frac{y}{2}\right)^2 + \left(r_0 \pm r\right)^2 + z_0^2} \equiv e^\pm_- \ , \\
d_{02}^\pm \stackrel{(\ref{VariableChange2})}{=} \sqrt{\left(y_0' + \frac{y}{2}\right)^2 + \left(r_0 \pm r\right)^2 + z_0^2} \equiv e^\pm_+ \ .
\end{split}
\end{align}

Thus
\begin{align} \label{ImpossibleIntegral}
\begin{split}
G_{1s_X}^{jj'}(x_1, x_2) &\stackrel{(\ref{VariableChange2})}{=} -\frac{\upsilon(\upsilon - 1)(X + 2)\lambda}{(2\pi)^3S}e^{is_X\theta_{12}}\delta^{jj'}H_{s_X}(r,y) \ , \\
H_{s_X}(r,y) &= \int_{\mathbb{R}^2}dy_0'dz_0\int_{0}^{\infty}dr_0\frac{1}{r_0}\frac{\left(4rr_0\right)^{2|s_X|}}{e_-^-e_-^+e_+^-e_+^+\left(e_-^- + e_-^+\right)^{2|s_X|}\left(e_+^- + e_+^+\right)^{2|s_X|}} \ .
\end{split}
\end{align}

The asymptotic of the integral $H_{s_X}(r,y)$ is carefully studied in \cite{1310.5078}.
\begin{align}
\begin{split}
G_{1s_X}^{jj'}(x_1, x_2) = \frac{\upsilon(\upsilon - 1)(X + 2)\e}{(X + 8)|s_X|}e^{is_X\theta_{12}}\delta^{jj'}\frac{\rho^{2(|s_X| + 1)}}{r^2}\log\rho + \mco(\rho^0) \ .
\end{split}
\end{align}

From (\ref{LogGreen'sFcnExp}) we know that we can find the first loop order correction to some of the CFT data from $(G_{0s_X}^{-1})^j{}_{j''}G_{1s_X}^{j''j}$, with $G_{0s_X}^{jj'}$ from (\ref{Green'sFCNSummand}). Taylor expanding $(G_{0s_X}^{-1})^j{}_{j''}$ around $\e = 0$
\begin{align}
\begin{split}
(G_{0s_X}^{-1})^j{}_{j''}G_{1s_X}^{j''j} = \frac{\upsilon(\upsilon - 1)(X + 2)\e}{2(X + 8)|s_X|}\delta^{jj'}\log\rho + \mco(\rho^0) + \mco(\e^2) \ .
\end{split}
\end{align}

Comparing this with the result from the OPE (\ref{LogOPE}) and we find that only $\Delta_{\psi_X}$ receives corrections from the one-loop diagram. This correction is given by
\begin{align}
\begin{split}
\Delta_{\psi_X}^1 = \frac{\upsilon(\upsilon - 1)(X + 2)\e}{2(X + 8)|s_X|} \ .
\end{split}
\end{align}

Putting it all together, up to one-loop corrections (or up to order $\e$), we have
\begin{align} \label{CFTDataConstraints}
\begin{split}
|c_X| &= \delta^{jj'} - \frac{\tilde{\psi}(|s_X| + 1) - \tilde{\psi}(1)}{4}\delta^{jj'}\e + \mco(\e^2) \ , \\
\Delta_{\psi_X} &= |s_X| + 1 - \left(1 - \frac{\upsilon(\upsilon - 1)(X + 2)}{(X + 8)|s_X|}\right)\frac{\e}{2} + \mco(\e^2) \ , \\
\Delta_{\phi_X} &= 1 - \frac{\e}{2} + \mco(\e^2) \ .
\end{split}
\end{align}

\begin{nt}
As a consistency check, one can see that this reduces to the results in \cite{1310.5078} when $X = 1$ and $\upsilon = 2^{-1}$
\begin{align}
\begin{split}
X = 1 \ , \quad \upsilon = \frac{1}{2} \quad\Rightarrow\quad \Delta_\psi = |s| + 1 - \left(\frac{1}{12|s|} + 1\right)\frac{\e}{2} + \mco(\e^2) \ .
\end{split}
\end{align}
\end{nt}

\section{Rychkov-Tan Analysis} \label{ChNoPertThy}

In this section we generalize the O($N$) framework created in \cite{1505.00963} to the WF O($N$) model with a co-dimension two, monodromy defect. This approach is very similar to that in \cite{1607.05551}. We define three axioms for the theory that contains information about its dynamics. This section will serve as another consistency check on the results (\ref{CFTDataConstraints}). In this section we need to rescale the bulk-local fields so that they matches the normalization (\ref{Normalization})
\begin{align} \label{RescaleBulk}
\begin{split}
\Phi^j \rightarrow \frac{1}{2\pi}\Phi^j \ .
\end{split}
\end{align}

\begin{ax}
The WF fixed point in the WF O($N$) model, see (\ref{FixedPoint}), is conformally invariant, hence the theory at this point is  a CFT.
\end{ax}

\begin{ax} \label{AxiomII}
Correlators in the WF fixed point approach free theory correlators (when the coupling constant is zero) in the limit
\begin{align}
\begin{split}
\e \rightarrow 0 \ .
\end{split}
\end{align}
This is because the coupling constant at this fixed point is proportional to $\e$. It yields that every operator in the $4-\e$ dimensional theory tends to operators in the free theory in the above limit.
\end{ax}

\begin{ax} \label{AxiomIII}
The operators
\begin{align} \label{Expr1}
\begin{split}
T_{2p} = \left(\phi_X^k\phi_X^k\right)^p \ , \quad T_{2p+1}^j = \phi_X^j\left(\phi_X^k\phi_X^k\right)^{p} \ , \quad j \ , k \in \{1,...,X\} \ ,
\end{split}
\end{align}
are all primary except $T_3^j$. The equations of motion from (\ref{Lagrangian}), with the rescaling of bulk-local operators (\ref{RescaleBulk}), tells us that it is a descendant of $T_1$
\begin{align} \label{Expr2}
\begin{split}
\alpha T_3^j = \p_\mu^2 T_1^j \ , \quad \alpha =  \frac{\lambda}{3!(2\pi)^2} = \frac{2\e}{X + 8} + \mathcal{O}(\e^2) \ .
\end{split}
\end{align}
\end{ax}

We will find $T_3^j$ first from (\ref{Expr1}) using Wick's theorem, and then compare it with the $T_3^j$ that we find from (\ref{Expr2}). The Wick contraction between two bulk-local primaries close to the defect is the propagator (\ref{RealProp}). From (\ref{Expr1}) we find
\begin{align} \label{AlmostW3Ax2}
\begin{split}
T_3^j = \frac{\upsilon(\upsilon - 1)(X + 2)}{2r^2}\phi_X^j + \mco(r^0) \ .
\end{split}
\end{align}

Using the bulk-defect expansion (\ref{BulkExpansion}) of $\phi^j_X$
\begin{align} \label{W3Ax2}
\begin{split}
T_3^j = \frac{\upsilon(\upsilon - 1)(X + 2)}{2}\sum_{s_X}\left(c_X\frac{e^{-is_X\theta}}{r^{\Delta_{\phi_X} - \Delta_{\psi_X} + 2}}\psi_{X}^{j} + \mco(r^{\Delta_{\phi_X} - \Delta_{\psi_X}}) \right) \ .
\end{split}
\end{align}

We move on to find $T_3^j$ using (\ref{Expr2}). With cylindrical coordinates
\begin{align*}
\begin{split}
T_3^j &= \alpha^{-1}\sum_{s_X}\left(c_X\left[\left(\Delta_{\phi_X} - \Delta_{\psi_X}\right)^2 - s_X^2\right]\frac{e^{-is_X\theta}}{r^{\Delta_{\phi_X} - \Delta_{\psi_X} + 2}}\psi_{X}^{j} + \mco(r^{\Delta_{\phi_X} - \Delta_{\psi_X}}) \right) \ .
\end{split}
\end{align*}

Compare the $r^{-\Delta_{\phi_X} + \Delta_{\psi_X} - 2}$-terms above with those in (\ref{W3Ax2}) to get the relation
\begin{align}
\begin{split}
\frac{\upsilon(\upsilon - 1)(X + 2)}{2} = \frac{\left(\Delta_{\phi_X} - \Delta_{\psi_X}\right)^2 - s_X^2}{\alpha} \ .
\end{split}
\end{align}

The scaling dimension, $\Delta_{\phi_X}$, for bulk-local operators is found using the framework for O($N$) models from \cite{1505.00963}. It is the same as in chapter \ref{ChGreenFcn}, see (\ref{CFTDataConstraints}). If we write $\Delta_{\psi_X}$ as a power series in $\e$, we find it to be the same as in chapter \ref{ChGreenFcn} as well.

\section*{Acknowledgement}

I want to thank Pietro Longhi and Joseph Minahan for introduction to the subject as well as interesting discussions in this project's early stages. I am grateful for the people who went to my presentation for this work and the institution of theoretical physics at Uppsala University for their hospitality. I want to thank Jian Qiu and Johan Henriksson for useful discussions, as well as Roberto Goranci and Seán Gray for help and support along the way.

\appendix

\section{Proper and Improper O($2$) Solutions}\label{AppEqSysSol}

In this appendix we solve the first two equations from (\ref{EquationSystem}) when $\sin\vartheta \ne 0$
\begin{align} \label{O(2)EqSys}
\begin{split}
\left\{ \begin{array}{l l}
e^{-2\pi is}C^1{}_{R,s}\Psi^R_s &= \pm\cos\vartheta C^1{}_{R,s}\Psi^R_s \mp\sin\vartheta C^2{}_{R,s}\Psi^R_s \ , \\
e^{-2\pi is}C^2{}_{R,s}\Psi^R_s &= \sin\vartheta C^1{}_{R,s}\Psi^R_s + \cos\vartheta C^2{}_{R,s}\Psi^R_s \ .
\end{array} \right.
\end{split}
\end{align}

The first of these equations yields
\begin{align} \label{C1C2Rel}
\begin{split}
C^1{}_{R,s}\Psi^R_s &= \mp\frac{\sin\vartheta}{e^{-2\pi is} \mp \cos\vartheta}C^2{}_{R,s}\Psi^R_s \ . \\
\end{split}
\end{align}

Inserting this into the second equation in (\ref{O(2)EqSys}) gives us
\begin{align} \label{EqToSolve}
\begin{split}
\left(e^{-2\pi is} - \cos\vartheta\right)\left(e^{-2\pi is} \mp \cos\vartheta\right) = \mp\sin^2\vartheta \ . \\
\end{split}
\end{align}

This will yield different results depending on whether $R_\vartheta$ in (\ref{UsedDefectDef}) has determinant one or minus one.

\subsection{Proper Rotation}

A proper $R_\vartheta$, i.e. $\det R_\vartheta = 1$, yields
\begin{align}
\begin{split}
\left(e^{-2\pi is} - \cos\vartheta\right)^2 &= -\sin^2\vartheta \ . \\
\end{split}
\end{align}

Solving for $s$
\begin{align}
\begin{split}
e^{-2\pi is} = \cos\vartheta \pm i\sin\vartheta = e^{\pm i(\vartheta + 2\pi n)} \ , \quad n\in\mathbb{Z} \quad\Leftrightarrow\quad s = n + \frac{\vartheta}{2\pi} \ .
\end{split}
\end{align}

Insert this back into (\ref{C1C2Rel}) and we find the relation
\begin{align}
\begin{split}
C^1{}_{R,s}\Psi^R_s &= \pm iC^2{}_{R,s}\Psi^R_s \ .
\end{split}
\end{align}

\subsection{Improper Rotation}

An improper $R_\vartheta$, i.e. $\det R_\vartheta = -1$, yields
\begin{align}
\begin{split}
\left(e^{-2\pi is} - \cos\vartheta\right)\left(e^{-2\pi is} + \cos\vartheta\right) &= \sin^2\vartheta \ . \\
\end{split}
\end{align}

Solving for $s$
\begin{align}
\begin{split}
e^{-4\pi is} = 1 \quad\Leftrightarrow\quad s = \frac{n}{2} \ , \quad n\in\mathbb{Z} \ .
\end{split}
\end{align}

\section{One-Loop Diagram Integral} \label{AppInt}

If we study the components of the integral (\ref{OneLoopGreen'sFcn}), we can solve it by carefully study its asymptotic expansion. First though, we need to massage the expression for the summand, $G_{0s_X}^{jj'}$, and then resum this expression to find the propagator $G_{0}^{jj'}$. The asymptotic behavior of (\ref{OneLoopGreen'sFcn}) will not be studied here. The interested reader may find details on its asymptotics in \cite{1310.5078}.

\subsection{Summand} \label{AppSummand}

We start with the summand $G_{0s}^{jj'}$. We cannot consider $r \equiv r_1 = r_2$, which corresponds to (\ref{Green'sFCNSummand}), since we are integrating over one of the coordinates. Thus we need to massage (\ref{PrimalZeroLoopGreen'sFcn}) using hypergeometric function relations
\begin{align}
\begin{split}
G_{0s}^{jj'}(x_k, x_l) &= \frac{\Gamma(|s| + 1)}{\Gamma(1)\Gamma(|s| + 1)}\frac{e^{is\theta_{kl}}}{r_kr_l}\alpha^{-(|s| + 1)}\delta^{jj'}\times \\
\eq \times{}_2F_1\left(|s| + 1, |s| + 1/2, 2|s| + 1, -4/\alpha\right) + \mco(\e) \\
&= \frac{e^{is\theta_{kl}}}{r_kr_l}\frac{4^s}{\sqrt{\alpha}\sqrt{4 + \alpha}\left(\sqrt{\alpha} + \sqrt{4 + \alpha}\right)^{2|s|}}\delta^{jj'} + \mco(\e) \ , \quad \alpha = 4\xi \ . 
\end{split}
\end{align}

Taylor expand this expression around $\e = 0$
\begin{align} \label{1stComp}
\begin{split}
G_{0s}^{jj'}(x_k, x_l) &= \frac{e^{is\theta_{kl}}}{r_kr_l}\frac{4^s}{\left(r_kr_l\right)^{-1}\sqrt{y_{kl}^2 + r_{kl}^2 + z_{kl}^2}\sqrt{4r_kr_l + y_{kl}^2 + r_{kl}^2 + z_{kl}^2}}\times \\
\eq \times\frac{1}{\left(r_kr_l\right)^{-s}\left(\sqrt{y_{kl}^2 + r_{kl}^2 + z_{kl}^2} + \sqrt{4r_kr_l + y_{kl}^2 + r_{kl}^2 + z_{kl}^2}\right)^{2|s|}}\delta^{jj'} + \\
\eq + \mco(\e) \\
&= e^{is\theta_{kl}}\frac{\left(4r_kr_l\right)^{|s|}}{d^-_{kl}d^+_{kl}\left(d^-_{kl} + d^+_{kl}\right)^{2|s|}}\delta^{jj'} + \mco(\e) \ , \\
d_{kl}^\pm &= \sqrt{y_{kl}^2 + (r_{kl}^\pm)^2 + z_{kl}^2} \ , \quad r_{kl}^\pm = r_k \pm r_l \ . 
\end{split}
\end{align}

\begin{nt}
The $z$-components are zero unless it is one of the integration variables in (\ref{OneLoopGreen'sFcn})
\begin{align}
\begin{split}
z_k = 0 \quad\text{if}\quad k \ne 0 \ . 
\end{split}
\end{align}
\end{nt}

\subsection{Resummation} \label{AppResum}

The next component in (\ref{OneLoopGreen'sFcn}) that we need to study is the sum $G_0^{jj'}(x_0,x_0)$. This component will be divergent, but we renormalize the theory such that we only care about its finite part. Let us denote
\begin{align}
\begin{split}
x \equiv \sqrt{y_{00}^2 + z_{00}^2} \quad\Rightarrow\quad d_{00}^- = \lim\limits_{x\rightarrow 0} x \ , \quad d_{00}^+ = \lim\limits_{x\rightarrow 0} \sqrt{(2r_0)^2 + x^2} \ .
\end{split}
\end{align}

We consider the defect-local operators in the bulk-defect expansion to have generic spin (\ref{Spin}) with $\upsilon$ fixed (since the operators we study in our Green's function transform in the same unbroken subgroup, O($X$), of O($N$)). Using (\ref{1stComp})
\begin{align} \label{Resum}
\begin{split}
G_0^{jj'}(x_0, x_0) &= \lim\limits_{x\rightarrow 0}\frac{\delta^{jj'}}{x\sqrt{(2r_0)^2 + x^2}}\sum_{s\in\mathbb{Z} + \upsilon}\left(\frac{2r_0}{\left(x + \sqrt{\left(2r_0\right)^2 + x^2}\right)}\right)^{2|s|} \ .
\end{split}
\end{align}

Resumming a geometric sum on the form
\begin{align}
\begin{split}
\sum_{s\in\mathbb{Z}+\upsilon}\eta^{|s|} = 2\sum_{s\geq\upsilon}\eta^{s} - \delta_{\upsilon 0} = \left[s' = s - \upsilon\right] = 2\sum_{s'\geq 0}\eta^{s' + \upsilon} - \delta_{\upsilon 0} = \frac{2\eta^\upsilon}{1 - \eta} - \delta_{\upsilon 0} \ ,
\end{split}
\end{align}

and using the following Taylor expansions 
\begin{align}
\begin{split}
\frac{1}{\sqrt{(2r_0)^2 + x^2}} &= \frac{1}{2r_0} + \mathcal{O}(x^2) \ ,
\end{split}
\end{align}
\begin{align}
\begin{split}
&\frac{1}{\sqrt{(2r_0)^2 + x^2}}\frac{\left(x + \sqrt{\left(2r_0\right)^2 + x^2}\right)^{-2\upsilon}}{1 - \left(2r_0\right)^2\left(x + \sqrt{\left(2r_0\right)^2 + x^2}\right)^{-2}} = \\
&\quad\quad\quad\quad\quad\quad\quad\quad\quad\quad  =  \frac{1}{\left(2r_0\right)^{2\upsilon}}\left(\frac{1}{2x} + \frac{1 - 2\upsilon}{2\left(2r_0\right)} + \frac{\upsilon(\upsilon - 1)}{\left(2r_0\right)^2}x\right) + \mathcal{O}(x^2) \ ,
\end{split}
\end{align}

yields
\begin{align} \label{2ndComp}
\begin{split}
G_0^{jj'}(x_0, x_0) &= \delta^{jj'}\left(\lim\limits_{x\rightarrow 0}\frac{1}{x}\left(\frac{1}{x} + \frac{1 - 2\upsilon - \delta_{\upsilon 0}}{2r_0}\right) + \frac{\upsilon(\upsilon - 1)}{2r_0^2}\right) \ .
\end{split}
\end{align}

We renormalize the theory such that we can ignore the divergent part ($x^{-2}$- and $x^{-1}$-terms) in the above propagator. This propagator is correct since we reproduce the result from \cite{1310.5078}, with an overall factor of $\delta^{jj'}$, in the half-integer case ($\upsilon = 1/2$), i.e.
\begin{align}
\begin{split}
G_0^{jj'}(x_0, x_0) &= \delta^{jj'}\left(\lim\limits_{x\rightarrow 0}\frac{1}{x^2} - \frac{1}{8r_0^2}\right) \ .
\end{split}
\end{align}

\bibliographystyle{utphys}
\footnotesize
\bibliography{ref}	

\providecommand{\href}[2]{#2}\begingroup\raggedright\begin{thebibliography}{10}

\bibitem{9711200}
J.~M. Maldacena, ``{The Large N limit of superconformal field theories and
  supergravity},'' \href{http://dx.doi.org/10.1023/A:1026654312961}{{\em Int.
  J. Theor. Phys.} {\bfseries 38} (1999) 1113--1133},
  \href{http://arxiv.org/abs/hep-th/9711200}{{\ttfamily arXiv:hep-th/9711200
  [hep-th]}}.
[Adv. Theor. Math. Phys.2,231(1998)].

\bibitem{0603001}
S.~Ryu and T.~Takayanagi, ``{Holographic derivation of entanglement entropy
  from AdS/CFT},'' \href{http://dx.doi.org/10.1103/PhysRevLett.96.181602}{{\em
  Phys. Rev. Lett.} {\bfseries 96} (2006) 181602},
\href{http://arxiv.org/abs/hep-th/0603001}{{\ttfamily arXiv:hep-th/0603001
  [hep-th]}}.

\bibitem{0807.0004}
R.~Rattazzi, V.~S. Rychkov, E.~Tonni, and A.~Vichi, ``{Bounding scalar operator
  dimensions in 4D CFT},''
  \href{http://dx.doi.org/10.1088/1126-6708/2008/12/031}{{\em JHEP} {\bfseries
  12} (2008) 031},
\href{http://arxiv.org/abs/0807.0004}{{\ttfamily arXiv:0807.0004 [hep-th]}}.

\bibitem{1212.3616}
A.~L. Fitzpatrick, J.~Kaplan, D.~Poland, and D.~Simmons-Duffin, ``{The Analytic
  Bootstrap and AdS Superhorizon Locality},''
  \href{http://dx.doi.org/10.1007/JHEP12(2013)004}{{\em JHEP} {\bfseries 12}
  (2013) 004},
\href{http://arxiv.org/abs/1212.3616}{{\ttfamily arXiv:1212.3616 [hep-th]}}.

\bibitem{1212.4103}
Z.~Komargodski and A.~Zhiboedov, ``{Convexity and Liberation at Large Spin},''
  \href{http://dx.doi.org/10.1007/JHEP11(2013)140}{{\em JHEP} {\bfseries 11}
  (2013) 140},
\href{http://arxiv.org/abs/1212.4103}{{\ttfamily arXiv:1212.4103 [hep-th]}}.

\bibitem{1611.01500}
L.~F. Alday, ``{Large Spin Perturbation Theory},''
\href{http://arxiv.org/abs/1611.01500}{{\ttfamily arXiv:1611.01500 [hep-th]}}.

\bibitem{1612.00696}
L.~F. Alday, ``{Solving CFTs with Weakly Broken Higher Spin Symmetry},''
\href{http://arxiv.org/abs/1612.00696}{{\ttfamily arXiv:1612.00696 [hep-th]}}.

\bibitem{1601.01784}
Y.~Hikida, ``{The masses of higher spin fields on $AdS_4$ and conformal
  perturbation theory},''
  \href{http://dx.doi.org/10.1103/PhysRevD.94.026004}{{\em Phys. Rev.}
  {\bfseries D94} no.~2, (2016) 026004},
\href{http://arxiv.org/abs/1601.01784}{{\ttfamily arXiv:1601.01784 [hep-th]}}.

\bibitem{1601.06794}
P.~Banerjee, S.~Datta, and R.~Sinha, ``{Higher-point conformal blocks and
  entanglement entropy in heavy states},''
  \href{http://dx.doi.org/10.1007/JHEP05(2016)127}{{\em JHEP} {\bfseries 05}
  (2016) 127},
\href{http://arxiv.org/abs/1601.06794}{{\ttfamily arXiv:1601.06794 [hep-th]}}.

\bibitem{1603.00387}
V.~Bashmakov, M.~Bertolini, L.~Di~Pietro, and H.~Raj, ``{Scalar Multiplet
  Recombination at Large N and Holography},''
  \href{http://dx.doi.org/10.1007/JHEP05(2016)183}{{\em JHEP} {\bfseries 05}
  (2016) 183},
\href{http://arxiv.org/abs/1603.00387}{{\ttfamily arXiv:1603.00387 [hep-th]}}.

\bibitem{1606.09593}
L.~F. Alday and A.~Bissi, ``{Unitarity and positivity constraints for CFT at
  large central charge},''
\href{http://arxiv.org/abs/1606.09593}{{\ttfamily arXiv:1606.09593 [hep-th]}}.

\bibitem{1610.06938}
A.~N. Manashov and E.~D. Skvortsov, ``{Higher-spin currents in the Gross-Neveu
  model at 1/n$^{2}$},'' \href{http://dx.doi.org/10.1007/JHEP01(2017)132}{{\em
  JHEP} {\bfseries 01} (2017) 132},
\href{http://arxiv.org/abs/1610.06938}{{\ttfamily arXiv:1610.06938 [hep-th]}}.

\bibitem{1611.10060}
D.~Mazac, ``{Analytic Bounds and Emergence of $\textrm{AdS}_2$ Physics from the
  Conformal Bootstrap},''
\href{http://arxiv.org/abs/1611.10060}{{\ttfamily arXiv:1611.10060 [hep-th]}}.

\bibitem{1701.04830}
P.~Liendo, ``{Revisiting the dilatation operator of the Wilson-Fisher fixed
  point},''
\href{http://arxiv.org/abs/1701.04830}{{\ttfamily arXiv:1701.04830 [hep-th]}}.

\bibitem{1703.03430}
C.~Behan, L.~Rastelli, S.~Rychkov, and B.~Zan, ``{Long-range critical exponents
  near the short-range crossover},''
\href{http://arxiv.org/abs/1703.03430}{{\ttfamily arXiv:1703.03430
  [cond-mat.stat-mech]}}.

\bibitem{1703.04830}
A.~Codello, M.~Safari, G.~P. Vacca, and O.~Zanusso, ``{Leading CFT constraints
  on multi-critical models in d>2},''
\href{http://arxiv.org/abs/1703.04830}{{\ttfamily arXiv:1703.04830 [hep-th]}}.

\bibitem{1703.05325}
C.~Behan, L.~Rastelli, S.~Rychkov, and B.~Zan, ``{A scaling theory for the
  long-range to short-range crossover and an infrared duality},''
\href{http://arxiv.org/abs/1703.05325}{{\ttfamily arXiv:1703.05325 [hep-th]}}.

\bibitem{1511.02921}
G.~Costagliola, ``{Operator product expansion coefficients of the 3D Ising
  model with a trapping potential},''
  \href{http://dx.doi.org/10.1103/PhysRevD.93.066008}{{\em Phys. Rev.}
  {\bfseries D93} no.~6, (2016) 066008},
\href{http://arxiv.org/abs/1511.02921}{{\ttfamily arXiv:1511.02921 [hep-th]}}.

\bibitem{1511.07108}
S.~M. Chester, M.~Mezei, S.~S. Pufu, and I.~Yaakov, ``{Monopole operators from
  the $4-\epsilon$ expansion},''
  \href{http://dx.doi.org/10.1007/JHEP12(2016)015}{{\em JHEP} {\bfseries 12}
  (2016) 015},
\href{http://arxiv.org/abs/1511.07108}{{\ttfamily arXiv:1511.07108 [hep-th]}}.

\bibitem{1512.00013}
M.~Hogervorst, S.~Rychkov, and B.~C. van Rees, ``{Unitarity violation at the
  Wilson-Fisher fixed point in 4-$\epsilon$ dimensions},''
  \href{http://dx.doi.org/10.1103/PhysRevD.93.125025}{{\em Phys. Rev.}
  {\bfseries D93} no.~12, (2016) 125025},
\href{http://arxiv.org/abs/1512.00013}{{\ttfamily arXiv:1512.00013 [hep-th]}}.

\bibitem{1603.04444}
Z.~Komargodski and D.~Simmons-Duffin, ``{The Random-Bond Ising Model in 2.01
  and 3 Dimensions},'' \href{http://dx.doi.org/10.1088/1751-8121/aa6087}{{\em
  J. Phys.} {\bfseries A50} no.~15, (2017) 154001},
\href{http://arxiv.org/abs/1603.04444}{{\ttfamily arXiv:1603.04444 [hep-th]}}.

\bibitem{1605.08087}
S.~El-Showk and M.~F. Paulos, ``{Extremal bootstrapping: go with the flow},''
\href{http://arxiv.org/abs/{1605.08087}}{{\ttfamily arXiv:{1605.08087}
  [hep-th]}}.

\bibitem{1510.07770}
K.~Sen and A.~Sinha, ``{On critical exponents without Feynman diagrams},''
  \href{http://dx.doi.org/10.1088/1751-8113/49/44/445401}{{\em J. Phys.}
  {\bfseries A49} no.~44, (2016) 445401},
\href{http://arxiv.org/abs/1510.07770}{{\ttfamily arXiv:1510.07770 [hep-th]}}.

\bibitem{1512.05994}
E.~D. Skvortsov, \href{http://dx.doi.org/10.1142/9789813144101_0008}{``{On
  (Un)Broken Higher-Spin Symmetry in Vector Models},''} in {\em {Proceedings,
  International Workshop on Higher Spin Gauge Theories: Singapore, Singapore,
  November 4-6, 2015}}, pp.~103--137.
\newblock 2017.
\newblock \href{http://arxiv.org/abs/1512.05994}{{\ttfamily arXiv:1512.05994
  [hep-th]}}.
\newblock
\url{https://inspirehep.net/record/1410620/files/arXiv:1512.05994.pdf}.
\newblock

\bibitem{1602.04928}
P.~Dey, A.~Kaviraj, and K.~Sen, ``{More on analytic bootstrap for O(N)
  models},'' \href{http://dx.doi.org/10.1007/JHEP06(2016)136}{{\em JHEP}
  {\bfseries 06} (2016) 136},
\href{http://arxiv.org/abs/1602.04928}{{\ttfamily arXiv:1602.04928 [hep-th]}}.

\bibitem{1607.07077}
Z.~Li and N.~Su, ``{Bootstrapping Mixed Correlators in the Five Dimensional
  Critical O(N) Models},''
\href{http://arxiv.org/abs/1607.07077}{{\ttfamily arXiv:1607.07077 [hep-th]}}.

\bibitem{1609.09820}
V.~Bashmakov, M.~Bertolini, and H.~Raj, ``{Broken current anomalous dimensions,
  conformal manifolds, and renormalization group flows},''
  \href{http://dx.doi.org/10.1103/PhysRevD.95.066011}{{\em Phys. Rev.}
  {\bfseries D95} no.~6, (2017) 066011},
\href{http://arxiv.org/abs/1609.09820}{{\ttfamily arXiv:1609.09820 [hep-th]}}.

\bibitem{1610.08472}
S.~Giombi, V.~Gurucharan, V.~Kirilin, S.~Prakash, and E.~Skvortsov, ``{On the
  Higher-Spin Spectrum in Large N Chern-Simons Vector Models},''
  \href{http://dx.doi.org/10.1007/JHEP01(2017)058}{{\em JHEP} {\bfseries 01}
  (2017) 058},
\href{http://arxiv.org/abs/1610.08472}{{\ttfamily arXiv:1610.08472 [hep-th]}}.

\bibitem{1612.05032}
P.~Dey, A.~Kaviraj, and A.~Sinha, ``{Mellin space bootstrap for global
  symmetry},''
\href{http://arxiv.org/abs/1612.05032}{{\ttfamily arXiv:1612.05032 [hep-th]}}.

\bibitem{1701.06997}
S.~Giombi, V.~Kirilin, and E.~Skvortsov, ``{Notes on Spinning Operators in
  Fermionic CFT},''
\href{http://arxiv.org/abs/1701.06997}{{\ttfamily arXiv:1701.06997 [hep-th]}}.

\bibitem{1702.03938}
F.~Gliozzi, A.~L. Guerrieri, A.~C. Petkou, and C.~Wen, ``{The analytic
  structure of conformal blocks and the generalized Wilson-Fisher fixed
  points},''
\href{http://arxiv.org/abs/1702.03938}{{\ttfamily arXiv:1702.03938 [hep-th]}}.

\bibitem{1601.02883}
M.~Billò, V.~Gonçalves, E.~Lauria, and M.~Meineri, ``{Defects in conformal
  field theory},'' \href{http://dx.doi.org/10.1007/JHEP04(2016)091}{{\em JHEP}
  {\bfseries 04} (2016) 091},
\href{http://arxiv.org/abs/1601.02883}{{\ttfamily arXiv:1601.02883 [hep-th]}}.

\bibitem{1602.06354}
A.~Gadde, ``{Conformal constraints on defects},''
\href{http://arxiv.org/abs/1602.06354}{{\ttfamily arXiv:1602.06354 [hep-th]}}.

\bibitem{1607.06155}
S.~Balakrishnan, S.~Dutta, and T.~Faulkner, ``{Gravitational dual of the
  R\'{e}nyi twist displacement operator},''
\href{http://arxiv.org/abs/1607.06155}{{\ttfamily arXiv:1607.06155 [hep-th]}}.

\bibitem{1608.05126}
P.~Liendo and C.~Meneghelli, ``{Bootstrap equations for $ \mathcal{N} $ = 4 SYM
  with defects},'' \href{http://dx.doi.org/10.1007/JHEP01(2017)122}{{\em JHEP}
  {\bfseries 01} (2017) 122},
\href{http://arxiv.org/abs/1608.05126}{{\ttfamily arXiv:1608.05126 [hep-th]}}.

\bibitem{1702.08471}
M.~Hogervorst and B.~C. van Rees, ``{Crossing Symmetry in Alpha Space},''
\href{http://arxiv.org/abs/1702.08471}{{\ttfamily arXiv:1702.08471 [hep-th]}}.

\bibitem{1210.4258}
P.~Liendo, L.~Rastelli, and B.~C. van Rees, ``{The Bootstrap Program for
  Boundary $CFT_d$},'' \href{http://dx.doi.org/10.1007/JHEP07(2013)113}{{\em
  JHEP} {\bfseries 07} (2013) 113},
\href{http://arxiv.org/abs/1210.4258}{{\ttfamily arXiv:1210.4258 [hep-th]}}.

\bibitem{1304.4110}
M.~Billó, M.~Caselle, D.~Gaiotto, F.~Gliozzi, M.~Meineri, and R.~Pellegrini,
  ``{Line defects in the 3d Ising model},''
  \href{http://dx.doi.org/10.1007/JHEP07(2013)055}{{\em JHEP} {\bfseries 07}
  (2013) 055},
\href{http://arxiv.org/abs/1304.4110}{{\ttfamily arXiv:1304.4110 [hep-th]}}.

\bibitem{1502.07217}
F.~Gliozzi, P.~Liendo, M.~Meineri, and A.~Rago, ``{Boundary and Interface CFTs
  from the Conformal Bootstrap},''
  \href{http://dx.doi.org/10.1007/JHEP05(2015)036}{{\em JHEP} {\bfseries 05}
  (2015) 036},
\href{http://arxiv.org/abs/1502.07217}{{\ttfamily arXiv:1502.07217 [hep-th]}}.

\bibitem{1605.04175}
F.~Gliozzi, ``{Truncatable bootstrap equations in algebraic form and critical
  surface exponents},'' \href{http://dx.doi.org/10.1007/JHEP10(2016)037}{{\em
  JHEP} {\bfseries 10} (2016) 037},
\href{http://arxiv.org/abs/1605.04175}{{\ttfamily arXiv:1605.04175 [hep-th]}}.

\bibitem{0303249}
O.~Aharony, O.~DeWolfe, D.~Z. Freedman, and A.~Karch, ``{Defect conformal field
  theory and locally localized gravity},''
  \href{http://dx.doi.org/10.1088/1126-6708/2003/07/030}{{\em JHEP} {\bfseries
  07} (2003) 030},
\href{http://arxiv.org/abs/hep-th/0303249}{{\ttfamily arXiv:hep-th/0303249
  [hep-th]}}.

\bibitem{1611.02485}
J.~Long, ``{On co-dimension two defect operators},''
\href{http://arxiv.org/abs/1611.02485}{{\ttfamily arXiv:1611.02485 [hep-th]}}.

\bibitem{1310.5078}
D.~Gaiotto, D.~Mazac, and M.~F. Paulos, ``{Bootstrapping the 3d Ising twist
  defect},'' \href{http://dx.doi.org/10.1007/JHEP03(2014)100}{{\em JHEP}
  {\bfseries 03} (2014) 100},
\href{http://arxiv.org/abs/1310.5078}{{\ttfamily arXiv:1310.5078 [hep-th]}}.

\bibitem{1505.00963}
S.~Rychkov and Z.~M. Tan, ``{The $\epsilon$-expansion from conformal field
  theory},'' \href{http://dx.doi.org/10.1088/1751-8113/48/29/29FT01}{{\em J.
  Phys.} {\bfseries A48} no.~29, (2015) 29FT01},
\href{http://arxiv.org/abs/1505.00963}{{\ttfamily arXiv:1505.00963 [hep-th]}}.

\bibitem{1506.06616}
P.~Basu and C.~Krishnan, ``{$\epsilon$-expansions near three dimensions from
  conformal field theory},''
  \href{http://dx.doi.org/10.1007/JHEP11(2015)040}{{\em JHEP} {\bfseries 11}
  (2015) 040},
\href{http://arxiv.org/abs/1506.06616}{{\ttfamily arXiv:1506.06616 [hep-th]}}.

\bibitem{1605.08868}
K.~Nii, ``{Classical equation of motion and Anomalous dimensions at leading
  order},'' \href{http://dx.doi.org/10.1007/JHEP07(2016)107}{{\em JHEP}
  {\bfseries 07} (2016) 107},
\href{http://arxiv.org/abs/1605.08868}{{\ttfamily arXiv:1605.08868 [hep-th]}}.

\bibitem{1612.08115}
K.~Roumpedakis, ``{Leading Order Anomalous Dimensions at the Wilson-Fisher
  Fixed Point from CFT},''
\href{http://arxiv.org/abs/1612.08115}{{\ttfamily arXiv:1612.08115 [hep-th]}}.

\bibitem{1510.04887}
S.~Ghosh, R.~K. Gupta, K.~Jaswin, and A.~A. Nizami, ``{$\epsilon$-Expansion in
  the Gross-Neveu model from conformal field theory},''
  \href{http://dx.doi.org/10.1007/JHEP03(2016)174}{{\em JHEP} {\bfseries 03}
  (2016) 174},
\href{http://arxiv.org/abs/1510.04887}{{\ttfamily arXiv:1510.04887 [hep-th]}}.

\bibitem{1510.05287}
A.~Raju, ``{$\epsilon$-Expansion in the Gross-Neveu CFT},''
  \href{http://dx.doi.org/10.1007/JHEP10(2016)097}{{\em JHEP} {\bfseries 10}
  (2016) 097},
\href{http://arxiv.org/abs/1510.05287}{{\ttfamily arXiv:1510.05287 [hep-th]}}.

\bibitem{1601.01310}
S.~Giombi and V.~Kirilin, ``{Anomalous dimensions in CFT with weakly broken
  higher spin symmetry},''
  \href{http://dx.doi.org/10.1007/JHEP11(2016)068}{{\em JHEP} {\bfseries 11}
  (2016) 068},
\href{http://arxiv.org/abs/1601.01310}{{\ttfamily arXiv:1601.01310 [hep-th]}}.

\bibitem{1609.00572}
R.~Gopakumar, A.~Kaviraj, K.~Sen, and A.~Sinha, ``{Conformal Bootstrap in
  Mellin Space},'' \href{http://dx.doi.org/10.1103/PhysRevLett.118.081601}{{\em
  Phys. Rev. Lett.} {\bfseries 118} no.~8, (2017) 081601},
\href{http://arxiv.org/abs/1609.00572}{{\ttfamily arXiv:1609.00572 [hep-th]}}.

\bibitem{1611.08407}
R.~Gopakumar, A.~Kaviraj, K.~Sen, and A.~Sinha, ``{A Mellin space approach to
  the conformal bootstrap},''
\href{http://arxiv.org/abs/1611.08407}{{\ttfamily arXiv:1611.08407 [hep-th]}}.

\bibitem{1611.06373}
C.~Hasegawa and {\relax Yu}.~Nakayama, ``{$\epsilon$-Expansion in Critical
  $\phi^3$-Theory on Real Projective Space from Conformal Field Theory},''
  \href{http://dx.doi.org/10.1142/S0217732317500456}{{\em Mod. Phys. Lett.}
  {\bfseries A32} no.~07, (2017) 1750045},
\href{http://arxiv.org/abs/1611.06373}{{\ttfamily arXiv:1611.06373 [hep-th]}}.

\bibitem{1611.10344}
F.~Gliozzi, A.~Guerrieri, A.~C. Petkou, and C.~Wen, ``{Generalized
  Wilson-Fisher Critical Points from the Conformal Operator Product
  Expansion},'' \href{http://dx.doi.org/10.1103/PhysRevLett.118.061601}{{\em
  Phys. Rev. Lett.} {\bfseries 118} no.~6, (2017) 061601},
\href{http://arxiv.org/abs/1611.10344}{{\ttfamily arXiv:1611.10344 [hep-th]}}.

\bibitem{1607.05551}
S.~Yamaguchi, ``{The $\epsilon$-expansion of the codimension two twist defect
  from conformal field theory},''
\href{http://arxiv.org/abs/1607.05551}{{\ttfamily arXiv:1607.05551 [hep-th]}}.

\bibitem{9505127}
D.~M. McAvity and H.~Osborn, ``{Conformal field theories near a boundary in
  general dimensions},''
  \href{http://dx.doi.org/10.1016/0550-3213(95)00476-9}{{\em Nucl. Phys.}
  {\bfseries B455} (1995) 522--576},
\href{http://arxiv.org/abs/cond-mat/9505127}{{\ttfamily arXiv:cond-mat/9505127
  [cond-mat]}}.

\bibitem{BookGroups}
H.~Weyl, {\em {The Classical Groups Their Invariants and Representations}}.
\newblock Princeton University Press; 2nd Revised ed. edition (October 13,
  1997), 1946.

\bibitem{1404.1094}
L.~Fei, S.~Giombi, and I.~R. Klebanov, ``{Critical $O(N)$ models in
  $6-\epsilon$ dimensions},''
  \href{http://dx.doi.org/10.1103/PhysRevD.90.025018}{{\em Phys. Rev.}
  {\bfseries D90} no.~2, (2014) 025018},
\href{http://arxiv.org/abs/1404.1094}{{\ttfamily arXiv:1404.1094 [hep-th]}}.

\end{thebibliography}\endgroup
	
\end{document}